\documentclass[12pt,a4paper]{article}
\usepackage{graphicx}
\usepackage{times}
\usepackage{natbib}
\usepackage{amssymb}
\usepackage{amsmath}

\begin{document}

%
%
%
%
\title{\bf Theory of Winds from Hot, Luminous Massive Stars}
%
%
%
%
\author{Stan Owocki$^1$\\
\normalsize $^1$ Department of Physics and
 Astronomy, University of Delaware, Newark, DE 19716 USA} 
%
\date{\mbox{}}
\maketitle
\pagestyle{empty}
%
%
\def\bull{\vrule height .9ex width .8ex depth -.1ex}
\makeatletter
\def\ps@plain{\let\@mkboth\gobbletwo
\def\@oddhead{}\def\@oddfoot{\hfil\tiny\bull\quad
``The multi-wavelength view of hot, massive stars''; 39$^{\rm th}$
Li\`ege Int.\ Astroph.\ Coll., 12-16 July 2010 \quad\bull}%
\def\@evenhead{}\let\@evenfoot\@oddfoot}
\makeatother
%
%
\def\beginrefer{\section*{References}%
\begin{quotation}\mbox{}\par}
\def\refer#1\par{{\setlength{\parindent}{-\leftmargin}\indent#1\par}}
\def\endrefer{\end{quotation}}
%
%
\vspace{-1cm}
{\noindent\small{\bf Abstract:} 
The high luminosities of massive stars drive strong stellar winds,
through line scattering of the star's continuum radiation.  
This paper reviews the dynamics of such line driving, building first upon the
standard CAK model for steady winds, and
deriving the associated analytic scalings for the mass loss rate and
wind velocity law.
It next summarizes the origin and nature of the strong Line-Deshadowing
Instability (LDI) intrinsic to such line-driving, including also the role of 
a diffuse-line-drag effect that stabilizes the wind base, and
then decribes how both instability and drag are incorporated in
the Smooth Source Function (SSF) method for time-dependent
simulations of the nonlinear evolution of the resulting wind structure.
The review concludes with a discussion of the effect of the resulting 
extensive structure in temperature, density and velocity for 
interpreting observational diagnostics.
In addition to the usual clumping effect on density-squared diagnostics, 
the spatial {\em porosity} of optically thick clumps can
reduce single-density continuum absorption, and a kind of
velocity porosity, or {\em vorocity}, can reduce the absorption strength
of spectral lines.
An overall goal is to illuminate the rich physics of
radiative driving and the challenges that lie ahead in developing
dynamical models for the often complex structure and variability of
hot-star winds.
}
%
%
\section{Introduction}
The strong stellar winds from hot, massive,
luminous stars are driven by the scattering of the star's continuum
radiation flux by line-transitions of metal ions
(Lucy \& Solomon 1970; Castor, Abbott \& Klein 1975, hereafter CAK).
The effectiveness of such line-driving depends crucially on the
Doppler-shifted line-desaturation arising from the wind outflow;  
this gives the dynamics of such winds an intricate feedback character,
in which the radiative driving force that accelerates the outflow
depends itself on that acceleration.
This leads to a strong, instrinsic Line-Deshadowing Instability (LDI) 
that is thought make such winds highly structured and variable.
The review here summarizes the basic dynamics of such line-driven
winds, with an emphasis on simulations of the nonlinear evolution of
instability-generated wind structure, and its implications for
interpreting wind diagnostics.


\section{The CAK/Sobolev Model for Steady Winds}

Consider a steady-state stellar wind outflow in which 
radiative acceleration $g_{rad}$ overcomes the 
local gravity $ GM_{\ast}/r^{2}$ at radius $r$ to drive a net acceleration
$v (dv/dr)$ in the radial flow speed $v(r)$. 
Since overcoming gravity is key, it is convenient to define a dimensionless 
equation of motion that scales all  accelerations by gravity,
\begin{equation}
(1-w_{s}/w) \, w' = -1 + \Gamma_{rad} 
\, ,
\label{dimlesseom}
\end{equation}
where $\Gamma_{rad} \equiv g_{rad} \, r^{2}/GM_{\ast}$,
$w \equiv v^{2}/v_{esc}^{2}$,
and $w' \equiv dw/dx$,
with  $x \equiv 1-R_{\ast}/r$ and $v_{esc} \equiv \sqrt{2GM_{\ast}/R_{\ast}}$
the escape speed from the stellar surface radius $R_{\ast}$. 
Eqn.\ (\ref{dimlesseom}) neglects gas pressure terms on the right side, 
since for isothermal sound speed $a$
these are of order $w_{s} \equiv (a/v_{esc})^{2} \approx 0.001$
compared to competing terms  needed to drive the wind.

For pure electron scattering opacity $\kappa_{e}$, the scaled radiative acceleration 
is just the usual Eddington parameter 
\begin{equation}
    \Gamma_{e} \equiv \frac{\kappa_{e} L_{\ast}}{4\pi GM_{\ast}c} 
     = 2 \times 10^{-5} \frac{L_{\ast}/L_{\odot}}{M_{\ast}/M_{\odot}} 
     \, .
\label{gamedddef}
\end{equation}
Because typically $L_{\ast} \sim M_{\ast}^{3}$, stars with 
$M_{\ast} > 10 M_{\odot}$ have $\Gamma_{e} > 10^{-3}$, with the
Eddington limit $\Gamma_{e} \rightarrow 1$ perhaps even being central
to setting a stellar upper mass limit of $M_{\ast} \sim 200 M_{\odot}$.
Eruptive mass loss from luminous blue
variable (LBV) stars like $\eta$~Carinae might in fact be continuum-driven during
episodes of super-Eddington luminosity 
(Davidson \& Humphreys 1997; Owocki, Gayley \& Shaviv 2004).

\begin{figure}
\includegraphics[width=14.5cm]{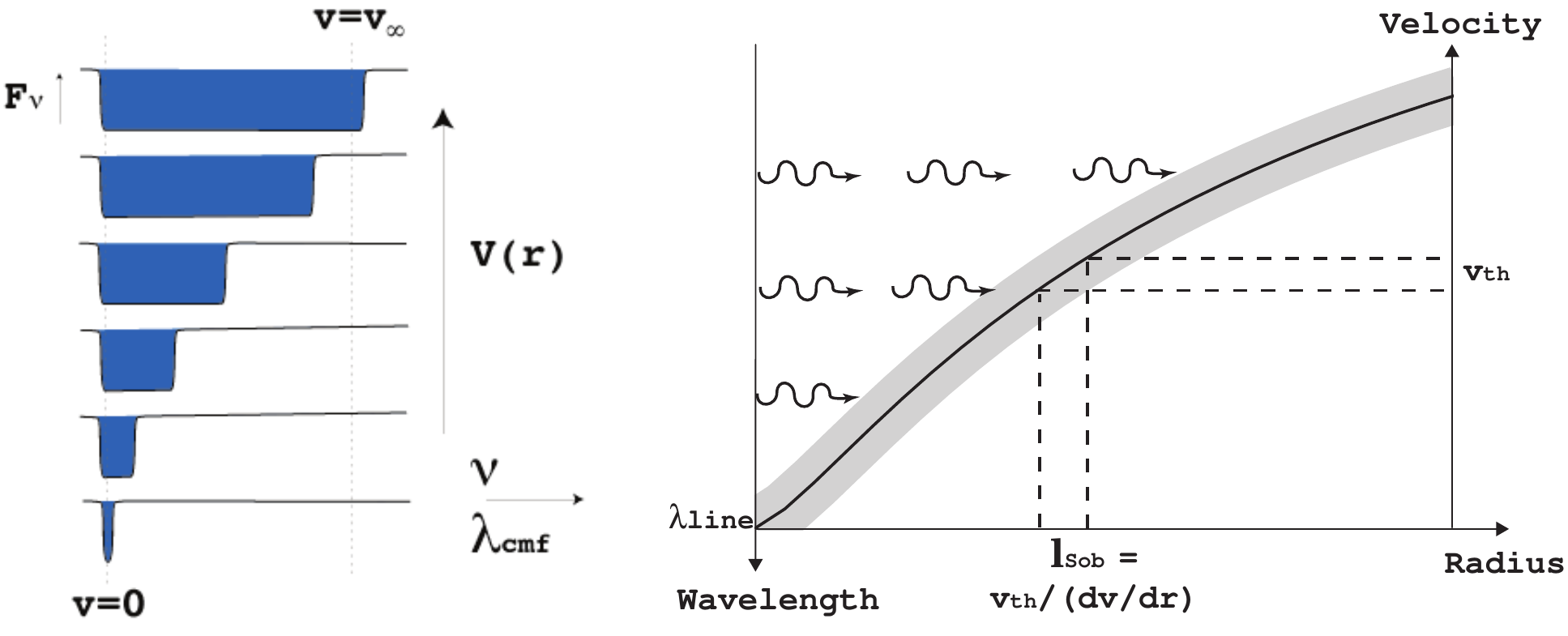}
\caption{
Two perspectives for 
the Doppler-shifted line-resonance in an accelerating flow. 
Right:
Photons with a wavelength just shortward of a line
propagate freely from the stellar surface up to a layer
where the wind outflow Doppler shifts the line into a resonance over a 
narrow width (represented here by the shading) equal to the Sobolev length,
set by the ratio of thermal speed to velocity gradient, 
$l_{Sob} \equiv v_{th}/(dv/dr)$. 
Left: 
Seen from successively larger radii within the accelerating wind, 
the Doppler-shift sweeps out an increasingly broadened line
absorption trough in the stellar spectrum.
}
\label{fig1}
\end{figure}

But the {\em resonant} nature of line (bound-bound) scattering from metal ions
leads to an opacity that is inherently much stronger than from free
electrons, by a factor set by the ``Quality'' $Q$ of the resonance, and
fraction $f_{b}$ of bound electrons in such metal lines (Gayley 1995).
For allowed transitions in the UV, $Q \sim 2 \times 10^{7}$, and for
solar metalicity, the total bound-electron fraction from all metal lines is 
$f_{b} \sim 10^{-4}$.
Thus,
in the somewhat idealized, {\it optically thin} limit that all 
the line opacity could be illuminated with a flat, unattenuated 
continuum from the full stellar luminosity, the total line-force 
would exceed the free-electron force by a factor of order 
${\overline Q} \sim Q f_{b} \sim 2000 $.
This implies line-driven winds can be initiated in 
even moderately massive stars with $\Gamma_{e} > 5 \times 10^{-4}$,
while for more massive stars with 
$\Gamma_{e} \approx 1/2$,
the net outward line acceleration in principle could be as 
high as $\Gamma_{thin} \approx {\overline Q} \Gamma_{e} \approx 1000$ 
times the acceleration of gravity!

In practice, self-absorption within strong lines limits the
acceleration, with the mass loss rate ${\dot M}$ set at the level 
for which the line driving is just sufficient to overcome gravity.
Indeed line-saturation keeps the dense, nearly static layers of the atmosphere
gravitationally bound.
But as illustrated by figure \ref{fig1}, within the accelerating wind, the Doppler shift 
of the line-resonance out of the absorption shadow of underlying material 
exposes the line opacity to a less attenuated flux.
This effectively desaturates the lines by limiting the resonance to a
layer with width set by the Sobolev length, $l_{Sob} = v_{th}/(dv/dr)$, 
and with optical depth proportional to
$t_{} \equiv \kappa_{e} \rho c /(dv/dr) $ 
$= \Gamma_{e} {\dot M} c^{2}/L_{\ast} w^{\prime}$.

For the CAK line-ensemble with a power-law number distribution in line-strength,
the cumulative force is reduced
by a factor  $1/({\overline Q}t)^{\alpha}$ from the optically thin 
value,
\begin{equation}
\Gamma_{CAK}
= \frac{ {\overline Q} \Gamma_{e} }{ (1-\alpha) ({\overline Q} t_{} )^{\alpha} } \, 
= \Gamma_{e} k t^{-\alpha} = C (w')^{\alpha} \, ,
\label {gamcak}
\end{equation}
where the second equality defines the CAK ``force muliplier''
$kt^{-\alpha}$, with\footnote{
Here we use a slight variation of the standard CAK notation in which
the artificial dependence on a fiducial ion thermal speed is avoided by
simply setting $v_{th} = c$.
Backconversion to CAK notation is achieved by multiplying
$t$ by $v_{th}/c$ and $k$ by $\left ( v_{th}/c \right)^{\alpha}$.
The line normalization ${\overline Q}$ 
offers the advantages of being a dimensionless measure of line-opacity 
that is independent of the assumed ion thermal
speed,  with a nearly constant characteristic value of order
${\overline Q} \sim 10^3$ for a wide range of ionization conditions
(Gayley 1995).}
$k \equiv {\overline Q}^{1-\alpha}/(1-\alpha)$.
The last equality relates the line-force to the flow acceleration,
with
\begin{equation}
C \equiv
\frac {1}{1-\alpha} \, \left [ \frac {L_{\ast}} {{\dot M} c^2} \right ]^\alpha \,
\left [ {\overline Q} \Gamma_{e} \right ]^{1-\alpha} \, .
\label {cdef}
\end{equation}
Note
that, for fixed sets of parameters for the star ($L_{\ast}$, $M_{\ast}$, 
$\Gamma_{e}$) and
line-opacity ($\alpha$, ${\overline Q}$), this constant scales with the mass loss rate 
as $C \propto 1/{\dot M}^{\alpha}$.

Neglecting the small sound-speed term $w_{s} \approx 0.001 \ll 1 $,
application of eqn.\ (\ref{gamcak}) into (\ref{dimlesseom}) gives 
the CAK equation of motion,
\begin{equation}
F = w' + 1 - \Gamma_{e} - C (w')^\alpha = 0 \, .
\label{cakeom}
\end{equation}
For small ${\dot M}$ (large $C$), there are two solutions, while for
large ${\dot M}$ (small $C$), there are no solutions.
The CAK critical solution corresponds to a {\it maximal} mass loss rate,
defined by $\partial F/\partial w' = 0$, for which the $C(w')^{\alpha}$ is
tangent to the line $1-\Gamma_{e} + w'$ at a critical acceleration
$w'_{c} = (1-\Gamma_{e}) \alpha/(1-\alpha )$.
Since the scaled equation of motion (\ref{cakeom}) has no explicit 
spatial dependence, this critical acceleration  applies throughout 
the wind, and so can be trivially integrated to yield 
$w(x) = w_{c}'\, x$.
In terms of dimensional quantities, this represents
a specific case of the general ``beta''-velocity-law,
\begin{equation}
v(r)=
v_\infty
\left ( 1- \frac {R_{\ast}}{r} \right )^{\beta} \, ,
\label{CAK-vlaw}
\end{equation}
where here $\beta=1/2$, and
the wind terminal speed $v_\infty = v_{esc} 
\sqrt{\alpha(1-\Gamma_{e})/(1-\alpha)}$.
Similarly, the critical value 
$C_{c}$
yields, through eqn.\ (\ref{cdef}), 
the standard CAK scaling for the mass loss rate
\begin{equation}
{\dot M}_{CAK}=\frac {L_{\ast}}{c^2} \; \frac {\alpha}{1-\alpha} \;
{\left[ \frac {{\overline Q} \Gamma_{e}}{1- \Gamma_{e}} \right]}^{{(1-\alpha)}/{\alpha}} \, .
\label{mdcak}
\end{equation}

These CAK results strictly apply only under the idealized assumption that
the stellar radiation is radially streaming from a point-source.
If one takes into account the finite angular extent of the stellar disk, 
then near the stellar surface the radiative force is reduced by 
a factor $f_{d\ast} \approx 1/(1+\alpha)$,
leading to a reduced mass loss rate 
(Friend \& Abbott 1986; Pauldrach, Puls \& Kudritzki 1986).
\begin{equation}
{\dot M}_{fd} = f_{d\ast}^{1/\alpha} {\dot M}_{CAK} 
= \frac{ {\dot M}_{CAK}}{(1+\alpha)^{1/\alpha}}
 \approx {\dot M}_{CAK}/2 \, .
\label{mdfd}
\end{equation}
Away from the star, the correction factor increases back toward unity,
which for the reduced base mass flux implies a stronger, more extended
acceleration, giving a somewhat higher terminal speed,
$v_{\infty} \approx 3 v_{esc}$, and a flatter velocity law,
approximated by replacing the exponent in
eqn.\ (\ref{CAK-vlaw}) by $\beta \approx 0.8$.

The effect of a radial change in ionization can be approximately taken into account
by correcting the CAK force (\ref{gamcak}) by a factor of the form
$
\left ( {n_e / W } \right )^\delta ,
$
where $n_e$ is the electron density, 
$W \equiv 0.5 ( 1-\sqrt{1-R_{\ast}/r} )$ 
is the radiation ``dilution factor'', and the exponent has a typical value 
$\delta \approx 0.1$ 
(Abbott 1982).
This factor introduces an additional density dependence to that already implied
by the optical depth factor $1/t_{}^\alpha$ given in eqn.\
(\ref{gamcak}).
Its overall effect can be roughly accounted with the simple
substition $\alpha \rightarrow \alpha' \equiv \alpha - \delta$ in the power
exponents of the CAK mass loss scaling law (\ref{mdcak}).
The general tendency is to moderately increase ${\dot M}$, and accordingly to
somewhat decrease the wind speed.

The above scalings also ignore the finite gas pressure associated with a
small but non-zero sound-speed parameter $w_{s}$.
Through a perturbation expansion of the equation of motion
(\ref{dimlesseom}) in this small parameter, it possible
to derive simple scalings for the fractional corrections 
to the mass loss rate and terminal speed
(Owocki \& ud-Doula 2004),
\begin{equation}
\delta m_{s} \approx \frac{ 4\sqrt{1-\alpha}} {\alpha} \, \frac{a}{ v_{esc}}
~~~~ ;  ~~~~ \delta v_{\infty,s} 
\approx \frac{ - \alpha \delta m_{s} }{ 2(1-\alpha) }
\approx \frac{-2 }{ \sqrt{1-\alpha} } \, \frac{a }{ v_{esc} }
\, .
\label{dmdvs}
\end{equation}
For a typical case with $\alpha \approx 2/3$ and $w_{s}=0.001$, 
the net effect is to increase the mass loss rate and decrease the wind 
terminal speed, both by about 10\%.

An important success of these CAK scaling laws is the theoretical
rationale
they provide for an empirically observed ``Wind-Momentum-Luminosity'' 
(WML) relation for OB supergiants 
(Kudrtizki, Lennon \& Pauldrach 1995).
Combining the CAK mass-loss law (\ref{mdcak}) together with the 
scaling of the terminal speed 
with the effective escape,
we obtain a WML relation of the form,
\begin{equation}
    {\dot M} v_{\infty} \sqrt{R_{\ast}} \sim L^{1/\alpha'} 
{\overline Q}^{1/\alpha'-1} 
\end{equation}
wherein we have neglected a residual dependence on $M(1-\Gamma_{e})$ that
is generally very weak for the usual case that $\alpha'$ is near $2/3$. 
Note that the direct dependence ${\overline Q} \sim Z$ provides the
scaling of the WML with metalicity $Z$.
Much current research aims also to understand deviations from this relation for
the weak winds of cooler OB dwarfs, and for the strong winds of
Wolf-Rayet stars (Puls et al. 2008).

\section{Non-Sobolev Models of Wind Instability}

The above CAK steady-state model depends crucially on the use of the
Sobolev approximation to compute the local CAK line force
(\ref{gamcak}).
Analyses that relax this approximation show that the flow is subject
to a strong, ``line-deshadowing instability'' (LDI) for 
velocity perturbations on a scale near and
below the Sobolev length $l_{Sob} = v_{th}/(dv/dr)$
(MacGregor, Harmann \& Raymond 1979; Owocki \& Rybick 1984, 1985).
Moreover, the diffuse, scattered component of the line force, 
which in the Sobolev limit is nullified by the fore-aft symmetry of
the Sobolev escape probability (see figure \ref{fig2}), 
turns out to have important dynamics effects on  the instability
through a ``diffuse line-drag"
(Lucy 1984).

\subsection{Linear Analysis of Line-Deshadowing Instability}

For sinusoidal perturbations ($\sim e^{i(kr-wt)}$) with wavenumber 
$k$ and frequency $\omega$, the linearized momentum equation
(ignoring the small gas pressure) relating
the perturbations in velocity and radiative acceleration implies
$
\omega = i \frac{\delta g}{\delta v}
$
,
which shows that unstable growth, with $\Im{\omega} > 0$, requires 
$\Re{(\delta g/\delta v}) > 0$.
For a purely Sobolev model 
(Abbott 1980),
the CAK scaling of the
line-force (\ref{gamcak}) with velocity gradient $v'$ implies
$\delta g \sim \delta v' \sim ik \delta v$, giving a purely real
$\omega$, and thus a stable wave that propagates inward at phase
speed,
\begin{equation}
\frac{\omega}{k} = - \frac{\partial g}{\partial v'} 
\equiv - U 
\, ,
\label{dgcak}
\end{equation}
which is now known as the ``Abbott speed''.
Abbott (1980) showed this is comparable to the outward wind flow speed, and
in fact exactly equals it at the CAK critical point.

\begin{figure}
    \includegraphics[width=17cm]{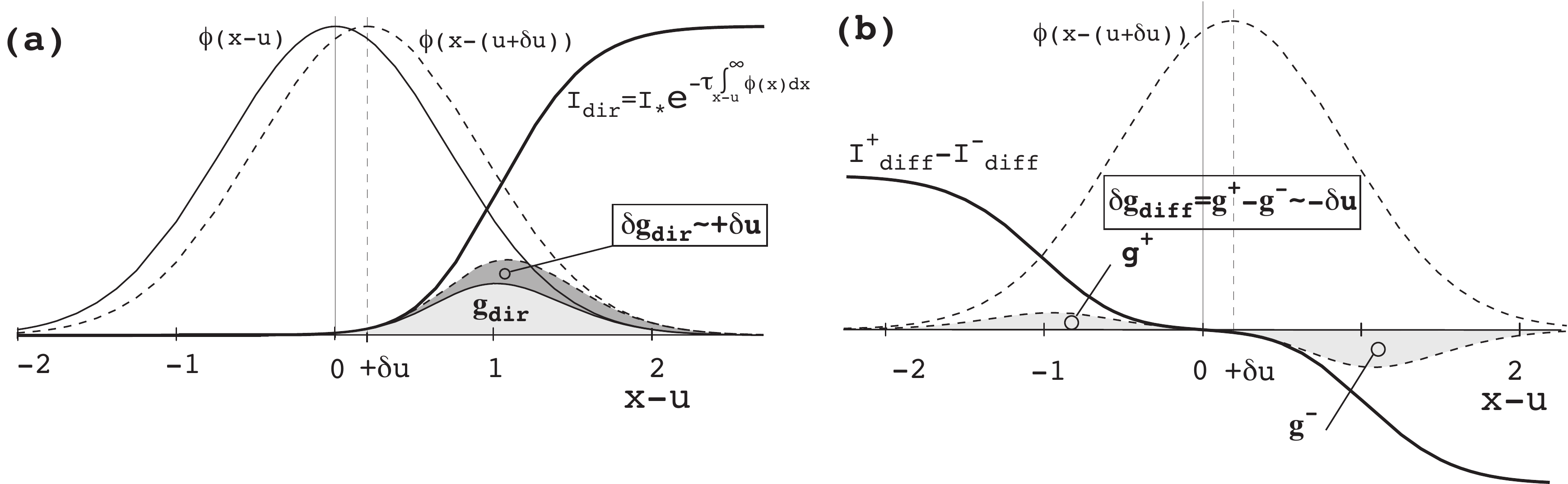}
\caption{
(a) The line profile $\phi$ and direct intensity plotted vs.
comoving frame frequency $x-u = x-v/v_{th}$, 
with the light shaded overlap area proportional 
to the net direct line-force $g_{dir}$.
The dashed profile shows the effect of the Doppler shift from a 
perturbed velocity $\delta v$, with 
the resulting extra area in the overlap with the blue-edge intensity 
giving a perturbed line-force $\delta g $ that scales in proportion to this
perturbed velocity $ \delta u = \delta v/v_{th}$.
(b) 
The comoving-frequency variation of the forward (+) 
and backward (-) streaming parts of the diffuse, scattered radiation.
Because of the Doppler shift from the perturbed velocity, 
the dashed profile has a stronger interaction with the backward 
streaming diffuse radiation, resulting in a diffuse-line-drag force 
that scales with the negative of the perturbed velocity, 
and so tends to counter the instability of the direct line-force in part a.
}
\label{fig2}
\end{figure}

As illustrated in figure \ref{fig2}a, instability arises from the deshadowing of
the line by the extra Doppler shift from the velocity perturbation,
giving $\delta g \sim \delta v$ and thus $\Im{\omega} > 0$.
A general analysis 
(Owocki \& Rybicki 1984)
yields a ``bridging law'' encompassing both
effects,
\begin{equation}
\frac{\delta g_{} }{ \delta v} 
\approx \Omega \frac { i k \Lambda }{  1 + ik \Lambda }
\, ,
\label{dgbridge}
\end{equation}
where 
$\Omega \approx g_{cak}/v_{th}$
sets the instability growth rate,
and the ``bridging length'' $\Lambda$ is found to be of order the 
Sobolev length $l_{sob}$.
In the long-wavelength limit $k \Lambda \ll 1$,
we recover the stable, Abbott-wave 
scalings of the Sobolev approximation,
$\delta g_{}/ \delta v  \approx i k \Omega \Lambda = ik U$;
while in the 
short-wavelength limit $k \Lambda \gg 1$,
we obtain the instability scaling
$\delta g_{}  \approx \Omega \delta v$.
The instability growth rate is very large, about the flow 
rate through the Sobolev length, $\Omega \approx v/l_{Sob}$.
Since this is a large factor $v/v_{th}$ bigger than the typical wind 
expansion rate $dv/dr \approx v/R_{\ast}$, a small perturbation 
at the wind base would, within this lineary theory,
be amplified by an enormous factor, of order 
$e^{v/v_{th}} \approx e^{100}$!

\subsection{Numerical Simulations of Instability-Generated Wind
Structure}

Numerical simulations of the nonlinear evolution require
a non-Sobolev line-force computation on a spatial grid that spans the 
full wind expansion over several $R_{\ast}$, yet resolves
the unstable structure at small scales near and below the Sobolev length.
The first tractable approach 
(Owocki, Castor \& Rybicki 1984)
focussed
on the {\em absorption} of the {\em direct} radiation from the stellar core,
accounting now for the attenuation from intervening material by
carrying out a {\em nonlocal integral} for the frequency-dependent
radial optical depth,
\begin{equation}
t(x,r) \equiv \int_{R_{\ast}}^r dr' \kappa_{e} \rho(r') 
\phi \left [ x-v(r')/v_{th} \right ]
\, ,
\label{txrad}
\end{equation}
where $\phi$ is the line-profile function, and
$x$ is the observer-frame frequency from line-center 
in units of the line thermal width.
The corresponding nonlocal form for the CAK line-ensemble force from 
this direct stellar radiation is
\begin{equation}
\Gamma_{dir} (r) = \Gamma_{e} {\overline Q}^{1-\alpha} 
\int_{-\infty}^{\infty} dx \, \frac{ \phi \left ( x-v(r)/v_{th} \right ) 
                           }{ t(x,r)^{\alpha} }
\, .
\label{gamdir}
\end{equation}
In the Sobolev approximation, $t(x,r) \approx \Phi (x-v/v_{th}) t$
(where $\Phi (x) \equiv \int_{x}^{\infty} \phi(x') \, dx'$),
this recovers the CAK form (\ref{gamcak}).
But for perturbations on a spatial scale near and below the 
Sobolev length, its variation also scales in proportion to the
perturbed velocity, leading to unstable amplification.
Simulations show that because of inward nature of wave propagation
implies an anti-correlation between velocity and density variation,
the nonlinear growth leads to high-speed rarefactions that steepen 
into strong {\em reverse} shocks and compress material into dense 
clumps (or shells in these 1D models) 
(Owocki et al. 1988).

The assumption of pure-absorption was criticized by Lucy (1984),
who pointed out that the interaction of a velocity perturbation with
the background, {\em diffuse} radiation from line-scattering 
results in a {\em line-drag} effect that reduces, and potentially
could even eliminate, the instability associated with the 
direct radiation from the underlying star.
The basic effect is illustrated in figure \ref{fig2}.
The fore-aft ($\pm$) symmetry of the diffuse
radiation leads to cancellation of the  $g_{+}$ and $g_{-}$
force components from the forward and backward streams, as
computed from a  line-profile with frequency centered on the local
comoving mean flow.
Panel b shows that the Doppler shift associated with the
velocity perturbation $\delta v$ breaks this symmetry, and leads to
stronger forces from the component opposing the perturbation.
 
Full linear stability analyses accounting for scattering effects 
(Owocki \& Rybicki 1985)
show the fraction of the direct instability that is canceled by the
line-drag  of the perturbed diffuse force depends on the ratio of the 
scattering source function $S$ to core intensity $I_{c}$,
\begin{equation}
s = \frac{r^{2}}{R_{\ast}^{2}} \, \frac{2S}{I_{c}} 
\approx \frac{1}{1 + \mu_{\ast}}
~~~ ; ~~~   \mu_{\ast} \equiv \sqrt{1-R_{\ast}^{2}/r^{2}}
\, ,
\label{sdef}
\end{equation}
where the latter approximation applies for the optically thin form
$2S/I_{c} = 1-\mu_{\ast}$.
The net instability growth rate thus becomes
\begin{equation}
\Omega (r) \approx \frac{g_{cak}}{v_{th}}  
\frac{ \mu_{\ast} (r) }{ 1+ \mu_{\ast} (r) }
\, .
\label{omnet}
\end{equation}
This vanishes near the stellar surface, where $\mu_{\ast}= 0$,
but it approaches half the pure-absorption rate far from the star, 
where $\mu_{\ast} \rightarrow 1$.
This implies that the outer wind is still very unstable, with
cumulative growth of ca.\ $v_{\infty}/2v_{th} \approx 50$ e-folds.

Most efforts to account for scattering line-drag in simulations of the
nonlinear evolution of the instability have centered on a 
{\em Smooth Source Function} (SSF) approach 
(Owocki 1991; Feldmeier 1995; Owocki \& Puls 1996, 1999).
This assumes that averaging over frequency and angle makes the
scattering source function relatively insensitive to flow
structure, implying it can be pulled out of the integral in the formal
solution for the diffuse intensity.
Within a simple {\em two-stream} treatment of the line-transport,
the net diffuse line-force then depends on the {\em difference} in the 
{\em nonlocal} escape probabilities $b_{\pm}$ associated with 
forward (+) vs.\ backward (-) {\em integrals} of the frequency-dependent
line-optical-depth (\ref{txrad}).
For a CAK line-ensemble, the net diffuse force can be written in a form 
quite analogous to the direct component (\ref{gamdir}),
\begin{equation}
\Gamma_{diff} (r) = \frac{\Gamma_{e} {\overline Q}^{1-\alpha}}
                          {2(1+\mu_{\ast})} \,
\left [ b_{-}(r) - b_{+}(r) \right ]
\, ,
\label{gamdiff}
\end{equation}
with
\begin{equation}
b_\pm (r) \equiv 
 \int_{-\infty}^\infty dx \,  
\frac{ \phi(x - v(r)/v_{th} ) }{ \left [ t_\pm (\pm x,r) \right ]^\alpha } \, 
\label{bpmdef}
\end{equation}
where
for $t_{-}$ the integral bounds in (\ref{txrad}) are now from 
$r$ to the outer radius $R_{max}$ 
(Owocki \& Puls 1996)
and
the overall normalization for $\Gamma_{diff}$ 
assumes the optically thin source function from eqn.\ (\ref{sdef}).
In the Sobolev approximation, both the forward and backward integrals 
give the same form, viz.\ 
$t_{\pm}(\pm x,r) \approx \Phi [\pm(x-v/v_{th})] t$,
leading to the net cancellation of the Sobolev diffuse force.
But for perturbations on a spatial scale near and below the 
Sobolev length, the perturbed velocity breaks the forward/back
symmetry (figure \ref{fig2}b), 
leading to perturbed diffuse force that
now scales in proportion to the {\em negative} of the perturbed velocity,
and thus giving the diffuse line-drag that reduces the
net instability by the factors given in (\ref{sdef}) and (\ref{omnet}).


\begin{figure}
\includegraphics[width=15cm]{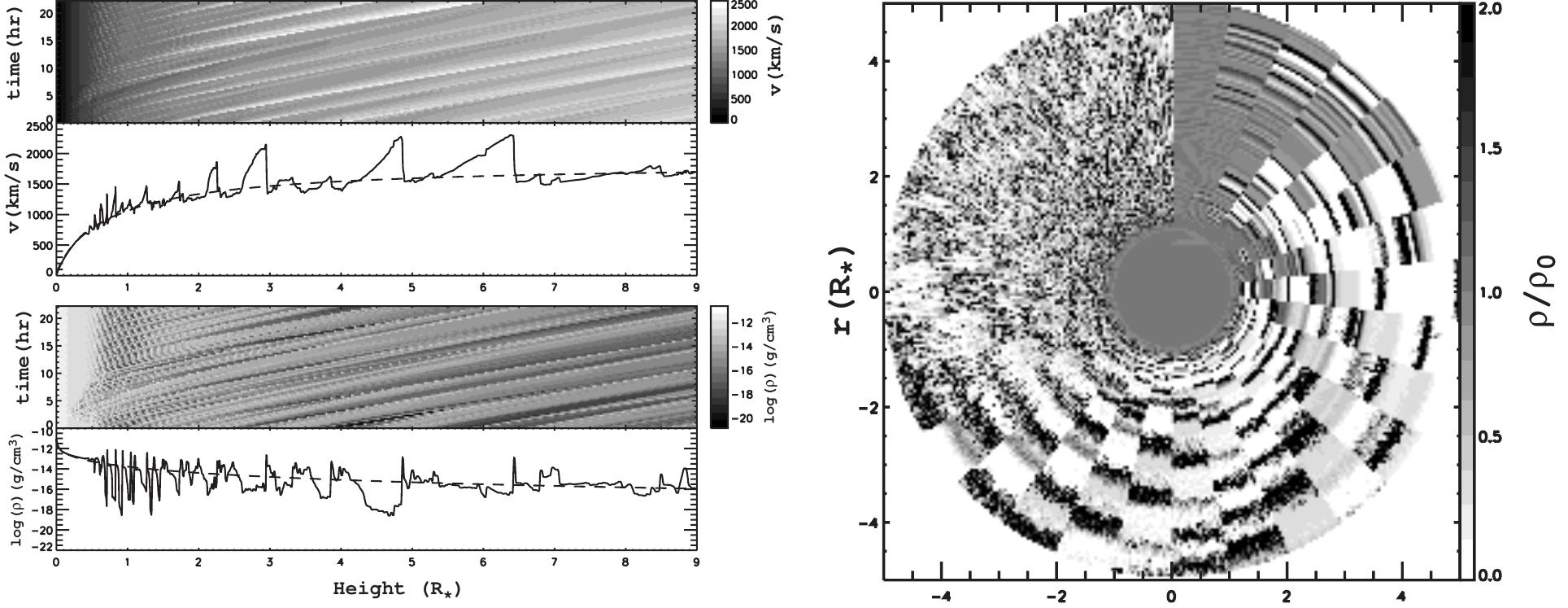}
\caption{
Left: Results of 1D Smooth-Source-Function (SSF) simulation of the 
line-deshadowing instability.
The line plots show the spatial variation of 
velocity (upper) and density (lower) at a fixed, arbitrary
time snapshot.
The corresponding grey scales show both the time (vertical axis)
and height (horizontal axis) evolution.
The dashed curve shows the corresponding smooth, steady CAK model.
Right: For 2DH+1DR SSF simulation, grayscale representation for the
density variations 
rendered as a time  sequence of 2-D wedges of the simulation model azimuthal 
range  $\Delta \phi = 12^{\rm o}$ stacked 
clockwise from the vertical in intervals of 4000~sec from the CAK
initial condition.
}
\label{fig3}
\end{figure}

The left panel of figure \ref{fig3} illustrates the results of a 1D SSF 
simulation, starting from an initial condition set by smooth, steady-state 
CAK/Sobolev model (dashed curves).
Because of the line-drag stabilization of the driving near the star 
(eqn. \ref{omnet}), the wind base remains smooth and steady.
But away from the stellar surface, the net strong instability leads to extensive 
structure in both velocity and density, roughly straddling the CAK steady-state.
Because of the backstreaming component of the diffuse line-force
causes any outer wind structure to induce small-amplitude fluctuations
near the wind base, the wind structure, once initiated, is  ``self-excited'', 
arising spontaneously without any explict perturbation from the
stellar boundary.

In the outer wind, the velocity variations become highly nonlinear and 
nonmonotonic, with amplitudes approaching $1000$~km/s, leading to 
formation of strong shocks. 
However, these high-velocity 
regions have very low density, and thus represent only very little material.
As noted for the pure-absorption models,
this anti-correlation between velocity and density arises because the 
unstable linear waves that lead to the structure have an {\it inward} 
propagation relative to the mean flow.
For most of the wind mass, the dominant overall effect of the instability 
is to concentrate material into dense clumps.
As discussed below, this 
can lead to
overestimates in the mass loss rate from
diagnostics 
that scale with the square of the density.


\begin{figure}
    \includegraphics[width=15cm]{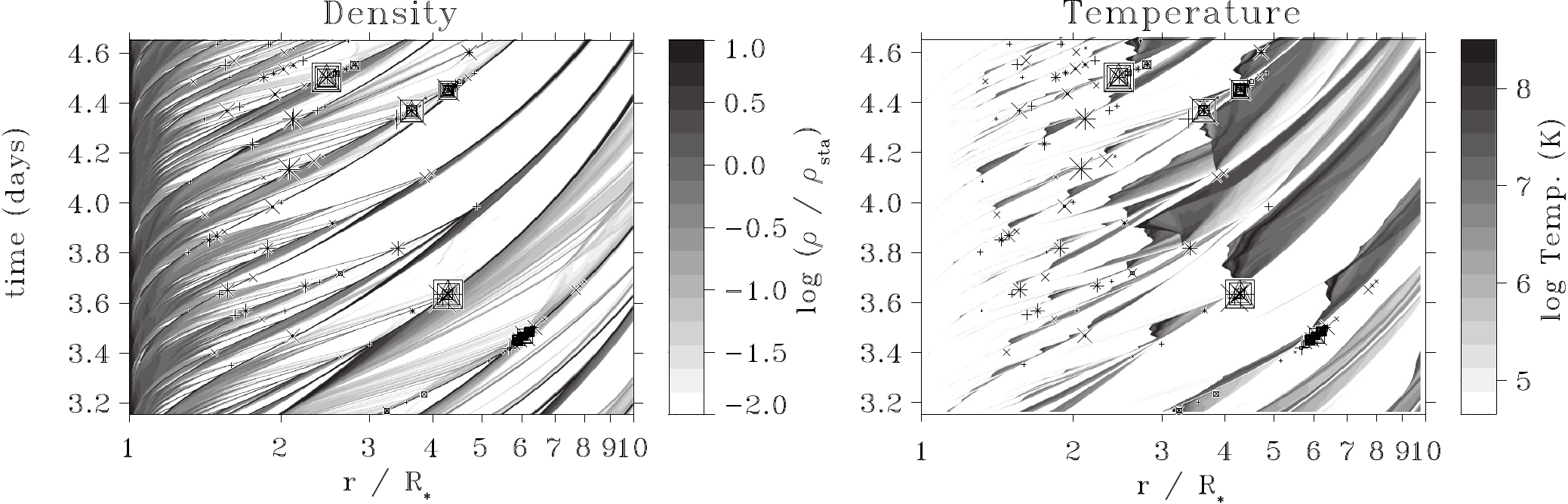}
\caption{
Greyscale rendition of the evolution of wind density and temperature,
for time-dependent wind-instability models with structure formation triggered 
by photospheric perturbations. 
The boxed crosses identify localized region of clump-clump collision that 
lead to the hot, dense gas needed for a substantial level of soft 
X-rays emission.
}
\label{fig4}
\end{figure}

The presence of multiple, embedded strong shocks suggests a potential 
source for the soft X-ray emission observed from massive star winds;
but the rarefied nature of the high-speed gas implies that this
self-excited structure actually
feeds very little material through 
the strong shocks needed to heat gas to X-ray emitting temperatures.
To increase the level of X-ray emission, 
Feldmeier, Pauldrach \& Puls (1997)
introduced intrinsic perturbations at the wind base, assuming 
the underlying stellar photosphere has a turbulent 
spectrum of compressible sound waves characterized by abrupt phase 
shifts in velocity and density.
These abrupt shifts seed wind variations that, when amplified by the 
line-deshadowing instabilty, now include substantial velocity 
variations among the dense clumps.
As illustrated in figure \ref{fig4}, when these dense clumps 
collide, they induce regions of relatively dense, hot gas which 
produce localized bursts of X-ray emission.
Averaged over time, these localized regions can 
collectively yield X-ray emission with a brightness and spectrum that 
is comparable to what is typically observed from such hot stars.
%

Because of the computational expense of carrying out nonlocal optical depth 
integrations at each time step, such SSF instability simulations have 
generally been limited to just 1D.
More realistically, various kinds of thin-shell instabilites 
(Visniac 1994)
can be expected to break up the structure 
into a complex, multidimensional form.
A first step to modelling both radial and lateral structure
(Dessart \& Owocki 2003)
is to use a restricted  ``2D-H+1D-R'' approach, extending the 
hydrodynamical model to 2D in radius and azimuth, 
but still keeping the  1D-SSF radial integration for the inward/outward 
optical depth within each azimuthal zone.
The right panel of figure \ref{fig3} shows
the resulting 2D density structure within a
narrow ($12^{\rm o}$) wedge, with the time evolution rendered 
clockwise at fixed time intervals of 4000 sec starting from the CAK
initial condition at the top.
The line-deshadowing instability is first manifest 
as  strong radial velocity variations and associated density compressions 
that initially extend nearly coherently across the full azimuthal range of 
the computational wedge.

But as these initial ``shell'' structures are accelerated outward, 
they become progressively  disrupted by Rayleigh-Taylor or thin-shell 
instabilities  that operate in azimuth down to the grid scale 
$d \phi = 0.2^{\rm o}$.
%
Such a 2DR+1DH approach may well exaggerate the level of variation on small 
lateral scales.
The lack of {\it lateral} integration needed to compute 
an azimuthal component of the diffuse line-force means that the model 
ignores a potentially strong net lateral line-drag that should 
strongly damp azimuthal velocity perturbations on scales below the lateral
Sobolev length $l_{0} \equiv r v_{th}/v_{r}$ 
(Rybicki, Owocki \& Castor 1990).
Presuming that this would inhibit development of lateral instability at 
such scales, then any lateral breakup would be limited to a minimum 
lateral angular scale of $\Delta \phi_{min} \approx l_{0}/r = 
v_{th}/v_{r} \approx 0.01 \, {\rm rad} \approx 0.5^{\rm o}$.
Further work is needed to address this issue through explicit 
incorporation of the lateral line-force and the associated line-drag 
effect.

\subsection{Clumping, Porosity and Vorosity: Implications for Mass Loss
Rates}

Both the 1D and 2D SSF simulations thus predict a wind with extensive 
structure in both velocity and density.
A key question then is how such structure might affect the various
wind diagnostics that are used to infer the mass loss rate.
Historically such wind clumping  has
been primarily considered for its effect on diagnostics that scale
with the square of the density,
The strength of such diagnostics is enhanced in a clumped wind,
leading to an overestimate of the wind mass loss rate that scales with
$\sqrt{f_{cl}}$, where the clumping factor
$f_{cl} \equiv \left < \rho^{2} \right >/
\left < \rho \right >^{2} $, with angle brackets denoting a local averaging
over many times the clump scale.
For strong density contrast between the clump and interclump medium,
this is just inverse of the clump volume filling factor, 
i.e. $f_{cl} \approx 1/f_{vol}$.
1D SSF simulations by Runacres \& Owocki (2002) generally find $f_{cl}$ increasing 
from unity at the structure onset radius $\sim 1.5 R_{\ast}$, peaking at a value
$f_{cl} \gtrsim 10$ at $r \approx 10 R_{\ast}$, with then a slow  outward decline to $\sim 5$
for $r \sim 100 R_{\ast}$.
These thus imply that thermal IR and radio emission formed in
the outer wind $r \approx 10-100 R_{\ast}$ may overestimate mass loss
rates by a factor 2-3.
The 2D models of Dessart \& Owocki (2004) find a similar variation,
but somewhat lower peak value,  and thus a lower
clumping factor than in 1D models, with a peak value of about 
$f_{cl} \approx 6$, apparently from the reduced collisional compression from 
clumps with different radial speeds now being able to pass by each  other.
But in both 1D and 2D models, the line-drag near the base means that
self-excited, intrinsic structure does not appear till $r \gtrsim 1.5 $, 
implying little or no clumping effect on $H \alpha$ line emission
formed in this region.
It should be stressed, however, that this is not necessarily a very
robust result, since turbulent perturbations at the wind base, and/or a
modestly reduced diffuse line-drag, might lead to onset of clumping
much closer to the wind base.

If clumps remain optically thin, then they have no effect on
single-density diagnostics, like the bound-free absorption of X-rays.
The recent analysis by Cohen et al. (2010) of the X-ray line-profiles 
observed by Chandra from $\zeta$-Pup indicates matching the relatively
modest skewing of the profile requires mass loss reduction of about a 
factor 3 from typical density-squared diagnostic value.
However, as discussed in the review by L. Oskinova in these
proceeedings, a key issue here is whether the individual clumps might
become {\em optically thick} to X-ray absorption.
In this case, the self-shadowing of material within the clump can the 
to an overall reduction in the effective opacity of the clumped medium
(Owocki, Gayley \& Shaviv 2004; Oskinova, Feldmeier, and Hamann 2007),
\begin{equation}
\kappa_{eff} = \kappa \frac{1 - e^{-\tau_{cl}}}{\tau_{cl}}
\, ,
\label{kapeffdef}
\end{equation}
where $\kappa$ is the microscopic opacity, and the
optical thickness for clumps of size $\ell$ is 
$\tau_{cl} = \kappa \rho \ell f_{cl}$.
The product $\ell f_{cl} \equiv h $ is known as the {\em porosity
length}, which also represents the {\em mean-free-path} between clumps.
A medium with optically thick clumps is thus porous, with an opacity reduction factor
$\kappa_{eff}/\kappa = 1/\tau_{cl} = 1/\kappa \rho h$.

However, it is important to emphasize that getting a significant
porosity decrease in the {\em continuum} absorption of a wind can be quite
difficult, since clumps must become optically thick near the radius of the
smoothed-wind photosphere, implying a collection of a
substantial volume of material into each clump, and so
a porosity length on order the local radius.
Owocki \& Cohen (2006) showed in fact that a substantial porosity reduction 
the absorption-induced asymmetry of X-ray line profiles required such 
large porosity lengths $h \sim r$.
Since the LDI operates on perturbations at the scale of the Sobolev
length $l_{sob} \equiv v_{th}/(dv/dr) \approx ({v_{th}/v_{\infty}})
R_{\ast} \approx R_{\ast}/300$,
the resulting structure is likewise very small scale,
as illustrated in the 2D SSF simulations in figure~\ref{fig3}.
Given the modest clumping factor $f_{cl} \lesssim 10$, it seems clear
that the porosity length is quite small, $h < 0.1 r$, 
and thus that porosity from LDI structure is not likely to be an important 
factor
\footnote{Okinova et al. 2004 argue that assuming clumps have a
flattened `pancake' shape with normal along the radial direction can allow 
greater transparency for X-rays passing near the star. 
But the above 2D SSF simulations suggest that Rayleigh-Taylor and thin-shell 
instabilities should break up such flattened pancakes into smaller, 
nearly spherical clumps.} 
for continuum processes like bound-free absorption of X-rays.

The situation is however quite different for {\em line} absorption,
which can readily be  optically thick in even a smooth wind, with
{\em Sobolev optical depth} 
$\tau_{sob} = \kappa_{l} \rho v_{th}/(dv/dr) = \kappa_{l} \rho
l_{sob} > 1$.
In a simple model with a smooth velocity law but material collected
into clumps with volume filling factor $f_{vol}=1/f_{cl}$, this clump 
optical depth would be even larger by a factor $f_{cl}$.
As noted by Oskinova, Hamman \& Feldmeier (2007),
the escape of radiation in the gaps between the thick clumps might then
substantially reduce the effective line strength, and so 
help explain the unexpected weakness of PV lines observed by FUSE 
(Fullerton et al. 2006), 
which otherwise might require a substantial, factor-ten or more reduction 
in wind mass loss rate.

\begin{figure*}[!t]
\begin{center}
\includegraphics
      [width=0.8\textwidth]{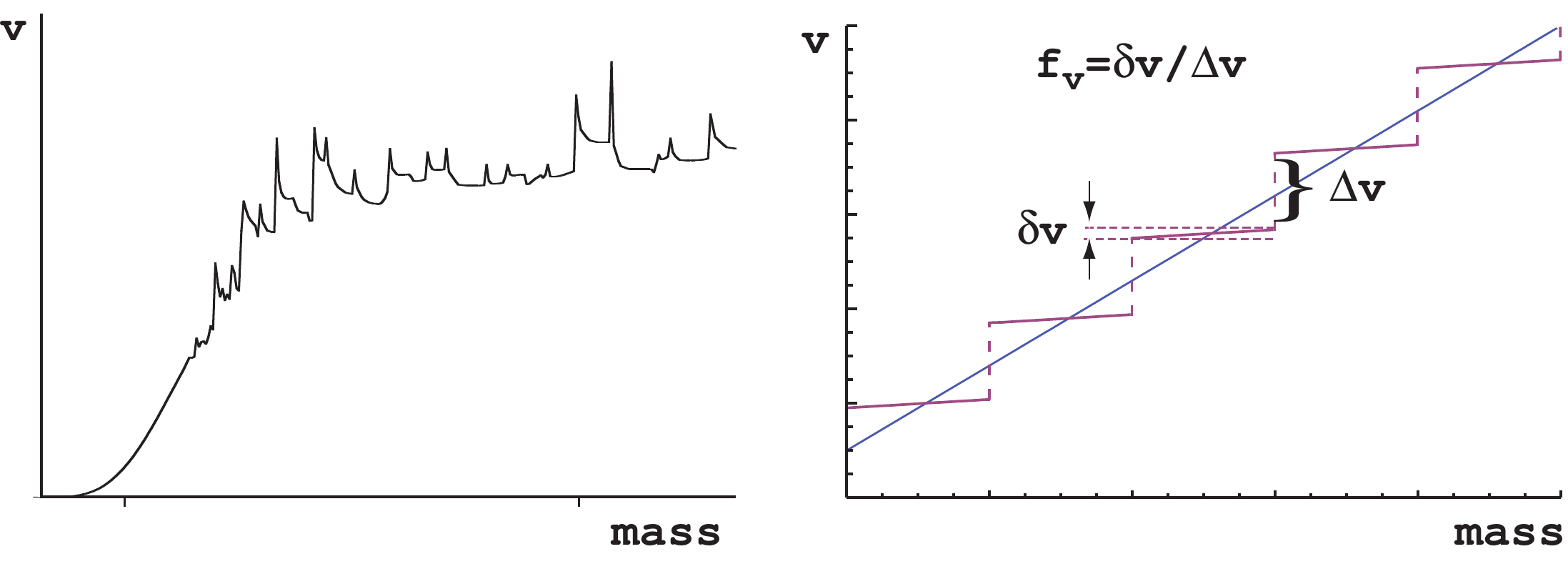}
\caption{
Left: Self-excited velocity structure arising in a 1D SSF simulation of the
line-driven instability, plotted versus a mass
coordinate, $M(r) = \int_{R}^{r} 4 \pi \rho r'^{2} \, dr' $.
Note the formation of velocity plateaus in the outer regions of the
wind.
Right: Velocity vs.\ mass in a wind seqment with structure described by
a simplified velocity staircase model with multiple large steps
$\Delta v$ between plateaus of width $\delta v$.
Here the associated velocity clumping factor 
$f_{vel} \equiv \delta v/\Delta v = 1/10 $.
The straight line represents the corresponding smooth 
CAK/Sobolev model.
}
\label{owocki-fig5}
\end{center}
\end{figure*}


But instead of {\em spatial} porosity, the effect on lines is better
characterized as a  kind of velocity porosity , or ``{\em vorosity}'',
which is now relatively insensitive to the spatial scale of wind structure
(Owocki 2008).
%
%
The left panel of figure \ref{owocki-fig5} illustrates the typical
result of 1D dynamical simulation of the wind instability, 
plotted here as a time-snapshot of velocity vs.\ a {\em mass} coordinate,
instead of radius.
The intrinsic instability of line-driving leads to a 
substantial velocity structure, with narrow peaks corresponding to spatially
extended, but tenuous regions of high-speed flow; these bracket dense, 
spatially narrow clumps/shells that appear here as nearly flat, extended
velocity plateaus in mass.
The right panel of figure \ref{owocki-fig5} illustrates a simplified, 
heuristic model of such wind structure for a representative wind section,
with the velocity clumping now represented by a simple ``staircase'' structure,
compressing the wind mass into discrete sections of the wind velocity law, 
while evacuating the regions in between;
the structure is characterized by a
``velocity clumping factor'' $f_{vel}$, set by the ratio between the 
internal velocity width $\delta v$ to the 
velocity separation $\Delta v$ of the clumps.
The straight line through the steps represents the
corresponding smooth wind flow.

\begin{figure*}[!t]
\begin{center}
\includegraphics
       [width=0.8\textwidth]{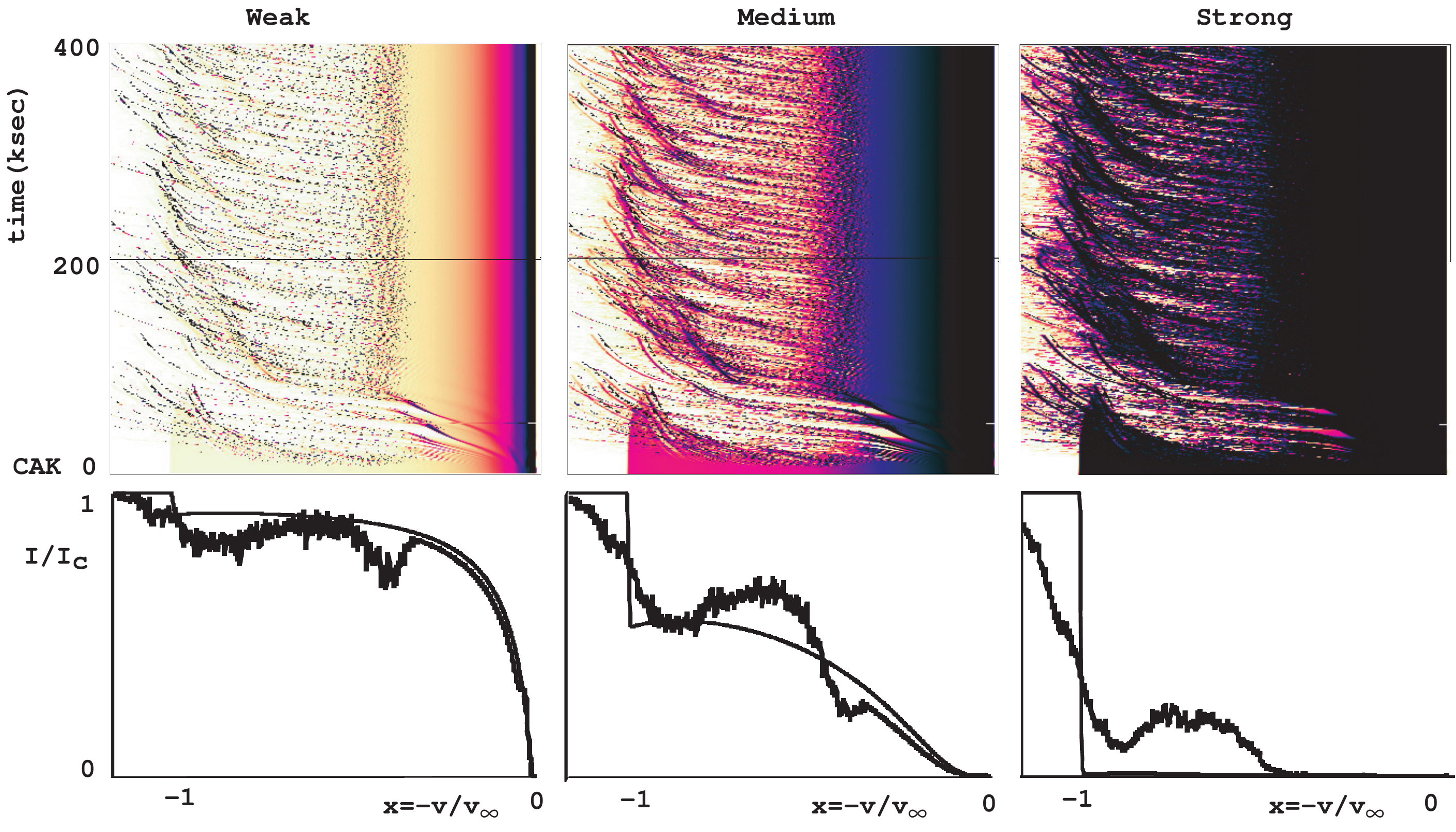}
\caption{
{\em Lower panels:}  Absorption trough of time-averaged P-Cygni line profile
plotted versus velocity-scaled observer wavelength $x=-v/v_{\infty}$
from line-center,
for a weak, medium, and strong line.
The smooth curves correspond to the smooth, CAK initial condition,
while the jagged curves represent results for 1D dynamical instability
simulations using the Smooth Source Function (SSF) method.
{\em Upper panels:} Color-scale plots of the associated dynamical
spectra, with time increasing vertically from the CAK initial
condition.
}
\label{owocki-fig6}
\end{center}
\end{figure*}

The effect of the velocity structure on the line-absorption  profile 
depends on the local Sobolev optical depth, 
which scales with the inverse of the mass derivative of velocity,
$\tau_{y} \sim 1/(dv/dm)$,
evaluated at a resonance location 
$r_{s}$, where  the velocity-scaled, observer-frame wavelength
$y =-v(r_{s})/v_{\infty}$.
In a smooth wind with Sobolev optical depth $\tau_{y}$, 
the absorption profile is given simply by (Owocki 2008),
$
A_{y} = 1 - e^{-\tau_{y}} .
$
In the structured model,
the optical thickness of individual clumps
is increased by the inverse of the clumping factor $1/f_{vel}$,
but they now only cover a fraction $f_{vel}$ of the
velocity/wavelength interval.
The net effect on the averaged line profile is to 
{\em reduce} the net aborption by a factor (Owocki 2008),
\begin{equation}
R_{A} (\tau_{y},f_{vel}) 
= f_{vel} ~ 
\frac{  1 - e^{-\tau_{y}/f_{vel}} }
      { 1 - e^{-\tau_{y}        } }
\, .
\label{owocki-radef}
\end{equation}
Note that for optically thick lines, $\tau_{y} \gg 1$, 
the reduction approaches a fixed value, given in fact by the clumping factor,
$R_{A} \approx f_{vel}$.
If the smooth-wind line is optically thin, $\tau_{y} \ll 1 $, then 
$R_{A} (\tau_{y}, f_{vel}) \approx (1 - e^{-\tau_{y}/f} )/
(\tau_{y}/f_{vel})$, which is quite analogous to the opacity reduction for
{\em continuum} porosity (eqn.\ \ref{kapeffdef}),
if we just substitute for the clump optical depth, 
$\tau_{c} \rightarrow \tau_{y}/f_{vel}$.

But a key point here is that, unlike for the continuum case, 
the {\em net reduction in line absorption no longer depends on
the spatial scale} of the clumps.
Instead one might think of this velocity clumping model as a kind 
of velocity form of the standard venetian blind, 
with $f_{vel}$ representing the fractional projected covering factor of the 
blinds relative to their separation.
The $f_{vel}=1$ case represents closed blinds that effectively block the
background light, while small $f_{vel}$ represent cases when the blinds
are broadly open, letting through much more light.

\subsection{Line-absorption profile from instability simulations}

%
%
Figure \ref{owocki-fig6} shows results for line-absorption spectra
from a typical 1D-SSF instability simulation,
wherein the intrinsic instabilty leads to extensive wind 
structure above a radius of about $r \approx 1.5 R_{\ast}$.
The upper panels of figure \ref{owocki-fig6} show the corresponding
effect on the dynamic spectra for a weak, mediuim, and strong line. 
The lower panels compare the associated time-average profile with that
of the smooth CAK initial condition.
%
The high level of velocity clumping 
leads to many tracks of enhanced, even saturated absorption, 
while at the same time exposing channels between the clumps that 
allow for increased tranmission of the stellar surface flux.
The time-averaged profiles in the lower panels thus show a general
{\em reduction} in the absorption compared to the smooth, CAK model,
most notably at middle wavelengths ($-y = v/v_{\infty} \approx 0.3-0.8$)
relative  to  blue edge for the CAK terminal speed $v_{\infty}$.
On the other hand, the unstable flow faster than the CAK
$v_{\infty}$ extends the absorption beyond $y=-1$, leading to notable 
softening of the blue edge.

But a key result here is that even the strong, saturated line has a residual 
flux of 10-20\%.
This is qualitatively just the kind of absorption reduction needed to 
explain the observed moderate strength of the PV line reported by 
Fullerton et al.\ (2006).
The contribution by Sundqvist in these proceedings describes recent
further efforts to account for unstable wind velocity structure in
quantitative modeling of both UV resonance lines like PV,
as well as recombination lines like $H \alpha $.
While demonstrating again the importance of accounting for velocity
and density structure for interpreting both diagnostics, the results
suggest a need for further development in radiation hydrodynamical
simulations to properly resolve the velocity structure of clumps, and
to induce their onset closer to the wind base, where $H \alpha$ is
formed.


   
\vspace{-0.25cm}

\footnotesize
\beginrefer

\vspace{-0.3cm}
\refer 
Abbott, D.~C.\ 1980, ApJ, 242, 1183.
    
\refer
Abbott, D.C.\ 1982, ApJ, 259, 282.

\refer
Castor, J., Abbott, D., \& Klein, R.\ 1975, ApJ 195, 157 (CAK).

\refer
Davidson, K., \& Humphreys, R.M.\ 1997, Ann Rev Astr Astrophys 35, 1  

\refer
Dessart, L. and Owocki, S.P.\ 2003, ApJ 406, 1.


\refer
Feldmeier, A.\ 1995, A\&A, 299, 523 

\refer
Feldmeier, A., Puls, J., and Pauldrach, A.\ 1997, A\&A 322, 878.


\refer
Friend, D.~B.~\& Abbott, D.~C.\ 1986, ApJ, 311, 701

\refer
Fullerton, A. W., Massa, D. L., \& Prinja, R. K.
2006, ApJ, 637, 1025

\refer
Gayley, K.\ 1995, ApJ 454, 410

\refer
Kudritzki, R.  Lennon, D., \& Puls, J.\ 1995, 
{\it Science with the VLT},  J.~Walsh \& I.~Danziger, 
eds., p. 246

\refer
Lucy, L.~B.\ 1984, ApJ, 284, 351 


\refer
Lucy, L. B., \& Solomon, P.\ 1970, ApJ 159, 879

\refer
MacGregor, K.~B., 
Hartmann, L., \& Raymond, J.~C.\ 1979, ApJ, 231, 514

\refer
Oskinova, L.~M., Feldmeier, A., \& Hamann, W.-R.
2006, MNRAS, 372, 313

\refer
Oskinova, L.~M., Hamann, W.-R., \& Feldmeier, A.\ 2007, A\&A 476, 1331 

\refer
Owocki, S.~P.\ 1991, NATO ASIC 
Proc.~341: Stellar Atmospheres - Beyond Classical Models, 235 


\refer
Owocki, S.~P., Castor,  J.~I., \& Rybicki, G.~B.\ 1988, 
ApJ 335, 914 

\refer
Owocki, S.~P., \& Cohen, D.~H.\ 2006, ApJ 648, 565 

\refer
Owocki, S., Gayley, K., \& Shaviv, N.\ 2004, ApJ 558, 802

\refer
Owocki, S.~P., \& Puls, J.\ 1996, ApJ 462, 894 

\refer
Owocki, S.~P., \& Puls, J.\ 1999, ApJ 510, 355 

\refer
Owocki, S.P. and Rybicki, G.B.\ 1984, ApJ 284, 337

\refer
Owocki, S.P. and Rybicki, G.B.\ 1985, ApJ 299, 265



\refer Owocki, S.~P.\ 2008, Clumping 
in Hot-Star Winds, 121,
W.-R. Hamann, A. Feldmeier \& L.M. Oskinova, eds.
Potsdam: Univ.-Verl., 2008
{\tt URN: http://nbn-resolving.de/urn:nbn:de:kobv:517-opus-13981}

\refer
Pauldrach, A., Puls, J., Kudritzki, R.P.\ 1986, A\&A 164, 86


%
%

%
\refer
Puls, J., Vink, J., \& Najarro, P. 2008, A\&A Reviews 16, 209

\refer
Rybicki, G.B., Owocki, S.P., and Castor, J.I.\ 1990, ApJ 349, 274



\refer
Sobolev, V.~V.\ 1960, {\it Moving Envelopes of Stars} 
(Cambridge: Harvard University Press).

\refer
Vishniac, E.T.\ 1994, ApJ 428, 186.

\endrefer


\end{document}